\documentclass[aps,amsmath,latexsym,amsfonts,floats,superscriptaddress]{JHEP3}
\usepackage{amsmath}
\usepackage{amsfonts}
\usepackage{amssymb}
\usepackage[all]{xy}

\newcommand{\qed}{\nobreak \ifvmode \relax \else
      \ifdim\lastskip<1.5em \hskip-\lastskip
      \hskip1.5em plus0em minus0.5em \fi \nobreak
      \vrule height0.75em width0.5em depth0.25em\fi}
   
\newcommand{\be}{\begin{equation}}
\newcommand{\ee}{\end{equation}}
\newcommand{\beq}{\begin{equation}}
\newcommand{\eeq}{\end{equation}}
\newcommand{\ba}{\begin{eqnarray}}
\newcommand{\ea}{\end{eqnarray}}
\newcommand{\bea}{\begin{eqnarray}}
\newcommand{\eea}{\end{eqnarray}}
\newcommand{\bean}{\begin{eqnarray*}}
\newcommand{\eean}{\end{eqnarray*}}
\newcommand{\bml}{\begin{mathletters}}
\newcommand{\eml}{\end{mathletters}}

\def\pp{/ \hspace {-2pt} /}

\def\beq{\begin{equation}}
\def\eeq{\end{equation}}

\def\E{{\cal{E}}}

\title{Self-gravitating branes of codimension 4 in Lovelock gravity}
\author{Robin Zegers~\footnote{email : robin.zegers@durham.ac.uk}\\
Department of Mathematical Sciences\\ University of Durham\\
South Road, Durham DH1 3LE, UK}
\abstract{We construct a familly of exact solutions of Lovelock equations describing codimension four branes with discrete symmetry in the transverse space. Unlike what is known from pure Einstein gravity, where such brane solutions of higher codimension are singular, the solutions we find, for the complete Lovelock theory, only present removable singularities. The latter account for a localised tension-like energy-momentum tensor on the brane, in analogy with the case of a codimension two self-gravitating cosmic string in pure Einstein gravity. However, the solutions we  discuss present two main distinctive features : the tension of the brane receives corrections from the induced curvature of the brane's worldsheet and, in a given Lovelock theory, the spectrum of possible values of the tension is discrete. These solutions provide a new framework for the study of higher codimension braneworlds.}
\preprint{DCPT-07/71}

\begin{document}



\vskip 1cm
By far the most widely studied braneworld scenario is the model in which the standard four dimensional universe is embedded as a hypersurface, {\it i.e.} a submanifold of codimension one, in a five dimensional space-time. However, higher codimension braneworlds have recently received a lot of attention, in particular because they exhibit new physical properties that may lead, for example, to interesting alternative approaches to the cosmological constant problem \cite{CV1}. Though it can be made sense of codimension two braneworlds, such models are usually very constrained in the framework of pure Einstein gravity. In particular, they can only support a tension-like matter content, the tension being related to the deficit angle of their conical transverse space. In higher codimensions, the situation is even worse and the only known solutions of Einstein equations are singular \cite{Ruth}. As pointed out in \cite{Greg}, Lovelock theory may naturally lead to a more realistic physics in codimension two braneworlds, by relaxing the constraint on the matter content of the brane. More recently, it has even been argued in \cite{CZ1, CZ2}, that, unlike Einstein equations, Lovelock equations could possibly admit self-gravitating brane solutions of codimension greater than two. However, no such solutions have been found up to now, apart from highly symmetric codimension two ones.

In this paper, we construct a new familly of codimension four self-gravitating branes, which provide a new framework for the study of codimension four braneworlds.  The construction relies on a distributional interpretation of the Euler form of quotient spaces admitting a minimal resolution of singularities. Such quotient spaces can be endowed with a familly of regularising metrics which, in some limit, tend to the singular metric of the brane's transverse space. It is in the sense of this limit, that the Euler form and, subsequently, the Lovelock field equations become distributions. Unlike the solutions of Einstein equations which present naked singularities, the only singularities of our solutions are therefore removable.

In the first section, we briefly review the basics of Lovelock theory and the relevant results of \cite{CZ1, CZ2} on matching conditions in higher codimensions. In the second section, we discuss the geometry of quotient spaces and the distributional interpretation of their Euler forms. In particular, we derive the distributional Euler forms of the quotients of ${\mathbb R}^4$ by the isometric action of the finite subgroups of $SO(4)$, making use of the corresponding self-dual instanton metrics as regularising metrics or, when direct computations get more involved, of standard results of algebraic geometry concerning resolutions of singularities. In the last section, we use these results to construct the promised codimension four brane solutions and discuss some of their main properties. In particular, we check that they solve the matching conditions derived in \cite{CZ1, CZ2} and provide the spectrum for the tension of these branes. We shall see that the value of the tension increases as the size of the transverse space is reduced by identifications under higher order groups, as we expect. However, unlike what is known from codimension two branes in pure Einstein gravity, where the tension can take on any real value, the spectrum of our codimension four branes is discrete. Moreover, for a given solution, the tension of the brane does not only depend on the volume of the transverse space but also on the induced curvature of the brane's worlsheet.

\section{Matching conditions in Lovelock gravity.}
\subsection{Lovelock gravity}
On a $D$-dimensional space-time $(M, g_D)$, Lovelock theory is the most general classical theory of gravity leading to second order field equations for the metric $g_D$ and to conservation of the energy-momentum tensor. The corresponding field equations, with energy-momentum tensor $T_{ab}$, read \cite{Lovelock}
\be
\label{love}
\sum_{k=0}^{\left[ \frac{D-1}{2} \right ]} \alpha_k {\mathcal E}_{(k)a} = -2T_{a}{}^b \theta^\star_{b} \, ,\ee
where the brackets stand for the integer part, the $\alpha_k$ are real
constants and the ${\mathcal E}_{(k)a}$ (with $a=1 \dots D$) are given by 
\be
{\mathcal E}_{(k)a} = \left (\bigwedge_{l=1}^k \Omega^{a_{2l-1} a_{2l}} \right
) \wedge \theta^\star_{a a_1 \cdots a_{2k}} \, ,
\ee
which is of order $k$ in the curvature 2-form
$\Omega^a{}_b=\frac{1}{2}R^a{}_{bcd} \theta^c \wedge \theta^d$. Finally,
$\theta^\star_{a_1 \cdots a_k} $ is the Hodge dual of $\theta^{a_1} \wedge
\cdots \wedge \theta^{a_k}$, the basis of the space of $k$-forms
$\Omega^{(k)}(TM)$, and we thus have
\be
\theta^\star_{a_1 \cdots a_k} = \frac{1}{(D-k)!} \epsilon_{a_1 \cdots a_k
  a_{k+1} \cdots a_D} \theta^{a_{k+1}} \wedge \cdots \wedge \theta^{a_D} \, .
\ee
When $D=4$, equation (\ref{love}) reduces to Einstein equations
($k=1$) with a cosmological constant $\alpha_0$, whilst for $D=5$, the
Gauss-Bonnet term ($k=2$) must be added and so on. Indeed, every two dimensions, a term of higher order in the curvature 2-form must be added to the field equations. The origin of these terms is better understood from the point of view of the Lovelock action
\be
S_L = \int_{M} \sum_{k=0}^{\left[ \frac{D-1}{2} \right ]} \alpha_k {\mathcal L}_{(k)} \, ,
\ee
where, for all $0 \leq k \leq \left[ (D-1)/2 \right ]$,
\be
 {\mathcal L}_{(k)} =  \left (\bigwedge_{l=1}^k \Omega^{a_{2l-1} a_{2l}} \right
) \wedge \theta^\star_{a_1 \cdots a_{2k}} \, ,
\ee
is simply the dimensional continuation of the Euler form in $2k$ dimensions. As emphasized in \cite{CZ1}, this topological origin of the theory is the reason of many of its particular properties, among which the fact that its field equations (\ref{love}) only involve derivatives of the metric up to second order and the possibility of considering self-gravitating sources of higher codimension. This latter feature, which we shall be interested in here, mathematically translates into the existence of non-trivial matching conditions for all values of the codimension.

\subsection{Matching conditions for arbitrary codimension}
Matching conditions are local requirements that the metric must satisfy in order for the field equations (\ref{love}) to admit a localised source, {\it i.e.} a distributional energy-momentum tensor of the form
\be
\label{set}
T_{ab}= \left ( \begin{array}{cc}
S_{\mu \nu} & 0 \\
0&0
\end{array} \right ) \delta_{\Sigma} \, ,
\ee
where $\delta_\Sigma$ denotes the Dirac distribution centered on the submanifold $\Sigma \hookrightarrow M$ and $S_{\mu\nu}$ is the tensor of energy-momentum density associated to the brane lying at $\Sigma$. Notice that the latter belongs to the tangential tensor space of the brane $T^{(2,0)}\Sigma$, which we indicate by using greek idices. From a physical point of view, the presence of the Dirac distribution $\delta_\Sigma$ means that the brane has no thickness, {\it i.e.} it has no extension in its transverse directions.

For codimension 1 and pure Einstein gravity, the matching conditions are given by the well-known 
Israel junction conditions, \cite{Israel} --see also \cite{darmois}--, where, if the induced 
metric is continuous,  a discontinuity in the first 
derivative of the
metric w.r.t the unique codimension accounts for the Dirac distribution on the l.h.s. of (\ref{love}) that precisely matches that on the r.h.s, due to (\ref{set}). These junction conditions generalise to Lovelock theory in all dimensions, \cite{JCLovelock} -- see also \cite{CZ1} for higher order contributions.

When the codimension is equal to 2, the simplest example of matching conditions is provided by the gravitational field of a straight cosmic string in four dimensions. In the far field approximation, it obeys Einstein equations with a distributional tension-like energy-momentum tensor localised on the string. The matching conditions, corresponding to the distributional part of the field equations, then relate the
tension of the string to the value $2\pi \beta$ of the overall conical defect it generates in the transverse plane \cite{Vilenkin}. This was first generalised to Einstein-Gauss-Bonnet gravity, {\it i.e.} for $0\leq k \leq 2$, in \cite{Greg}, yielding modified matching conditions that depend not only on the tension of the brane but also on its induced curvature through its induced Einstein tensor. As a consequence, the energy-momentum tensor of the brane is no longer restricted to a tension but can be, at least in principle, arbitrary. Like finite width corrections \cite{bonjour}, the Gauss-Bonnet term therefore allows non-maximally symmetric embeddings for the brane. Although it has been proven that there exist no isotropic cosmological solutions for braneworlds of codimension two with axial symmetry in the bulk and conserved energy-momentum tensor on the brane \cite{Kofinas}, this result is of major interest in the context of cosmological braneworld scenarios since it allows {\it a priori} more realistic energy-momentum tensors and induced geometries on codimension two braneworlds.

As proven in \cite{Geroch}, in pure Einstein gravity, under certain regularity conditions, one can only make sense of localised sources of codimension less than two. From a mathematical point of view, the field equations do not admit a distributional interpretation otherwise and the general solutions of Einstein equations with the corresponding symmetries will usually present non-removable singularities \cite{Ruth}. One way around this result, in higher dimensional space-times, comes from the non-trivial contribution of the higher order Lovelock terms. In \cite{CZ1,CZ2}, the matching conditions for self-gravitating branes of arbitrary codimensions were derived in the context of Lovelock gravity. In the particular case of $p$-branes of even codimension $N=2n$, such that 
\be
\label{isa}
\left[ p/2\right] +1 \leq n \, ,
\ee 
{\it i.e.} such that the codimension of the brane is greater than its intrinsinc dimension, it was shown that the matching conditions just reduce to the induced Lovelock equations reduced on $\Sigma$. For all $P \in \Sigma$, we thus have
\be
\label{topo}
 \sum_{\tilde{k}=0}^{\left [{p\over 2}\right ]} 
\tilde{\alpha}_{\tilde{k}} \E^{\Sigma}_{(\tilde{k})\mu}(P) = -2 S_\mu^\nu (P)
\theta^{\star {\pp}}_\nu \, ,
\ee
where
\be
\label{simple}
\E^{\Sigma}_{(\tilde{k})\mu}= \left (\bigwedge_{l=0}^{\tilde{k}} 
\Omega^{\lambda_{l} \nu_{l}}_{\Sigma} \right ) \wedge 
\theta^{\star {\pp}}_{\mu \lambda_1 \nu_{1} \cdots \lambda_{\tilde{k}} \nu_{\tilde{k}}} \in \Omega^{(p)}(T\Sigma)
\ee
denotes the induced Lovelock form of order $\tilde{k}$ and we have set
\be
\label{alphatilde}
\tilde{\alpha}_{\tilde{k}}=2^{2n-1} (\chi-\beta) \mbox{Area}({\mathbb S}_{2n-1}) \frac{(\tilde{k}+n)!(n-1)!}{\tilde{k}!}  \alpha_{\tilde{k}+n} \, .
\ee
Like in the cosmic string case, the quantity $\beta \mbox{Area}({\mathbb S}_{2n-1})$ appearing in (\ref{alphatilde}) measures the solid angle deficit generated by the presence of the brane in its transverse space and $\chi$ denotes the Euler characteristic of a geodesic ball in the transverse space of the brane, which is just $1$ for the two-dimensional disk in the case of the cosmic string. Notice that from (\ref{alphatilde}), we deduce that the Lovelock contribution of order $\tilde{k}$ to (\ref{topo}) on $\Sigma$ comes from the term of order $\tilde{k}+n=k$ in the field equations (\ref{love}).

In the next sections, we propose an original construction of self-gravitating branes of codimension four. It is based on a distributional interpretation of the Euler form of four dimensional cones built on quotient spaces of the 3-sphere ${\mathbb S}^3$ by the free action of finite subgroups of its isometry group $SO(4)$. As we shall see, these solutions will satisfy the matching conditions given in (\ref{topo}).

\section{A distributional interpretation for the Euler form}
\subsection{The Euler form of regularisable conical spaces}
In this section, we consider the geometric properties of conical spaces of even dimension, in particular with regards to the definition of their Euler form. Let $(M,g_M)$ be a riemannian manifold of even dimension $2n$, such that $M={\mathbb R}^*_+ \times N \cup \{ 0_M\}$ and 
\be
\label{regg0}
g_M = dr^2 + r^2 g_N \, ,
\ee
where $(N,g_N)$ is a $2n-1$ dimensional riemannian manifold. Let $\Gamma \subset \mbox{Isom} (N,g_N)$ be a finite subgroup of its isometry group with free action on $N$. The quotient space $N/\Gamma$ is a riemannian manifold and there exists a continuous surjection $\phi : N \rightarrow N/\Gamma$, such that $\phi \circ \gamma = \phi$ for all $\gamma \in \Gamma$. The metric $g_N$ on $N$ also naturally induces a metric $g_{N/\Gamma}$ on the quotient space according to $g_N = \phi^* g_{N/\Gamma}$. We can now define the conical space $M/\Gamma$ with base space $N/\Gamma$, as the $2n$-dimensional topological space ${\mathcal C}_{2n}=M/\Gamma = {\mathbb R}^*_+ \times N/\Gamma \cup \{0_M \}$, endowed with the metric
\be
\label{g0}
g_0 = dr^2 + r^2 g_{N/\Gamma} \, .
\ee
Notice that the notation $M/\Gamma$ is meaningful. It really denotes a quotient space since the free isometric action of $\Gamma$ on each sheet $N$ of $M$ induces an isometric action of $\Gamma$ on $M$ which is free everywhere on $M$ except at $0_M$. Unlike the metric of $M$ (\ref{regg0}), the metric (\ref{g0}) is regular everywhere, except possibly at $r=0$, {\it i.e.} at the fixed point $0_M$ of the action of $\Gamma$ on $M$, where it may be non-differentiable. We will assume that this conical space is regularisable, if it admits a continuous family of everywhere differentiable metrics $(g(t))_{t \in [0,1]}$ such that 
\be
\lim_{t \to 0} g(t) = g_0 \, .
\ee
In the examples we will consider, such regularisations of the metric will arise from a minimal resolution $\pi : \widehat{M/\Gamma} \rightarrow M/\Gamma$ of the singular quotient $M/\Gamma$ by a non-singular manifold $\widehat{M/\Gamma}$, the size of the exceptional divisor $\pi^{-1}(0_M)$ being controlled by $t$.
For all values of the real parameter $t$, one can define a torsion free metric connection 1-form $\omega^a{}_b(t)$ as well as the associated curvature 2-form $\Omega^a{}_b(t)$. A natural representative of the Euler class of $\widehat{{\mathcal C}_{2n}}$ is then
\be
E_{2n}(t) = \frac{1}{(2n)!} \epsilon_{a_1 a_2 \cdots a_{2n}} \Omega^{a_1a_2}(t) \wedge \cdots \wedge \Omega^{a_{2n-1}a_2n}(t) \, .
\ee 
Notice that, in general, the limit of this quantity as $t$ tends to zero is singular at $r=0$ since, there, the metric $g_0$ is not differentiable. However we can make sense of this limit as a distribution. For all ${\mathcal C}^1$-function $f :U\rightarrow {\mathbb R} $ with compact support in $U \subset \widehat{{\mathcal C}_{2n}}$, we thus define the distribution $\hat{E}_{2n}$ as
\be
\left\langle \hat{E}_{2n} ; f \right\rangle_U = \lim_{t \to 0} \int_U f E_{2n}(t) \, .
\ee
From this definition, we see that if $0_M$ is not contained in the interior $\mathring{U}$ of $U$, we just recover the integral of the smooth limit $E_{2n}(0)|_{U}$; whereas if $0_M \in \mathring{U}$, we should expect, in the limit, an extra contribution to the integral which can be interpreted as $E_{2n}(0)|_{U}$ being a Dirac distribution centered at $0_M$. If we assume that $U$ has a compact boundary $\partial U$, then, for all $t \in (0,1]$, the Chern-Gauss-Bonnet theorem \cite{Chern} holds for $U$, {\it i.e.}
\be
\int_U E_{2n}(t) = \left ( 4\pi \right )^{n} n!\, \chi(U) - \int_{\partial U} {\bf n}^* \Phi_{2n} (t) \, ,
\ee
where ${\bf n}$ is the unit outward pointing normal to $\partial U$ and the Chern-Simons form $\Phi_{2n}(t)$ can be expressed in terms of the connection 1-form $\omega^{a}{}_b(t)$ and of its curvature 2-form $\Omega^a{}_b(t)$ as
\ba
\Phi_{2n}(t) &=& \sum_{l=0}^{n-1} \frac{2n \cdots 2(l+1)}{1 \cdots (2(n-l)-1)}  \epsilon_{a_1 \cdots a_{2n-1}} \Omega^{a_1 a_2}(t) \wedge \cdots \wedge \Omega^{a_{2l-1} a_{2l}}(t) \wedge \omega^{a_{2l+1}}{}_{n}(t) \wedge \cdots \nonumber \\
&& \qquad \qquad\qquad\qquad \qquad\qquad \qquad \wedge \omega^{a_{2n-1}}{}_{n}(t) \, .
\ea
In the limit, we thus get a distributional version of the Chern-Gauss-Bonnet theorem, namely
\be
\label{DistCGB}
\left\langle \hat{E}_{2n} ; 1 \right\rangle_U = \left ( 4\pi \right )^{n} n! \, \chi(U) - \int_{\partial U} {\bf n}^* \Phi_{2n} (0) \, .
\ee
Notice that, since $\chi(U)$ is a topological invariant, its value is left unchanged when we vary $t$. Furthermore, the limit of the integral of the Chern-Simons form on the boundary $\partial U$ is regular as long as $0_M$ is not contained in it and is therefore equal to the integral of the Chern-Simons form for $g_0$. In particular, if we take $U$ to be the geodesic ball of radius $R$ in the metric $g_0$ (\ref{g0}), its boundary $\partial U$ is isometric to $N/\Gamma$ and we thus have
\be
\label{boundary}
\int_{\partial U} {\bf n}^* \Phi_{2n} (0) = \int_{N/\Gamma} {\bf n}^* \Phi_{2n} (0) =  \frac{1}{|\Gamma |} \int_{N} {\bf n}^* \Phi_{2n} (0) \, ,
\ee
where $| \Gamma |$ denotes the order of the finite subgroup $\Gamma$.
In practical applications, when $\chi$ can be obtained otherwise, this provides a simple way to determine the normalisation of the distributional Euler form.
We will show explicit examples of this in four dimensions, based on some specific cases of three dimensional spherical geometries, as base spaces.

\subsection{Spherical 3-manifolds}
In the next section, we will study the geometry of space-times describing 3-branes of codimension four. As we shall see, the transverse space of these objects is a four dimensional conical space with base space ${\mathbb S}^3 /\Gamma$, where $\Gamma \subset SO(4)$ is a finite subgroup of the isometry group $SO(4)$ of the 3-sphere, with free action on it. Fortunately, all such actions have been classified \cite{Seifert1, Seifert2} and the resulting quotient spaces are thus known. In this subsection, we briefly review these results, emphasizing the features that we shall need later on.

A remarkable property of ${\mathbb S}^3$ is that it is homomorphic to the non-abelian group $SU(2)$, or equivalently to the quaternion algebra. This can be seen by embedding the 3-sphere into ${\mathbb C}^2$, {\it i.e.}
\be
{\mathbb S}^3 = \left\lbrace (z_1, z_2) \in {\mathbb C}^2 \, , |z_1|^2+ |z_2|^2 = 1 \right\rbrace \, .
\ee
The mapping with $SU(2)$ reads
\be
{\mathbb S}^3 \ni (z_1, z_2) \rightarrow \left ( \begin{array} {cc} z_1 & -\bar{z}_2 \\ 
                                          z_2 & \bar{z}_1  \end{array} \right ) \in SU(2) \, .
\ee
This mapping induces a group structure on the 3-sphere with left and right actions on itself. The isometry group $SO(4)$ of the 3-sphere thus factorises as $SU_L(2) \times SU_R(2)$, which, in turn, allows us to describe the action of all its finite subgroups as different actions of the well-known finite subgroups of $SU(2)$. The classification of the finite subgroups of $SO(4)$ was achieved in \cite{Seifert1, Seifert2}. Restricting to free actions, it reads as follows
\begin{itemize}
\item{single left action of ${\mathbb Z}_n$, $D_m^*$, $T^*$, $O^*$ and $I^*$;}
\item{double action of 
\begin{itemize}
\item{$ {\mathbb Z}_m \times {\mathbb Z}_n$, with gcd($m$,$n$)$=1$;} 
\item{$D_{m}^* \times {\mathbb Z}_n$, with $m \geq 2$ and gcd($4m$,$n$)$=1$;} 
\item{$T^* \times {\mathbb Z}_n$, with gcd(24,$n$)$=1$;}
\item{$O^* \times {\mathbb Z}_n $, with gcd(48,$n$)$=1$;} 
\item{$I^* \times {\mathbb Z}_n $, with gcd(120,$n$)$=1$;}
\end{itemize}
}
\item{linked action of 
\begin{itemize}
\item{${\mathbb Z}_m \times_{\ell} {\mathbb Z}_n$, with $0 < n< m$ and gcd($m$,$n$)$=1$;}
\item{$D_m^* \times_{\ell} {\mathbb Z}_{8n}$, with gcd($m$,$8n$)$=1$;}
\item{$T^* \times_{\ell} {\mathbb Z}_{9n}$, with $n$ odd;}
\end{itemize}
} 
\end{itemize}
where $D^*_{m}$, $T^*$, $O^*$ and $I^*$ respectively denote the double covers of the dihedral, tetrahedral, octahedral and icosahedral finite subgroups of $SO(3)$. The double and linked actions concern the same group products but, in the linked action case, each element in the right action subgroup is paired with a restricted set of elements of the left action subgroup, taking care to avoid fixed points on the 3-sphere. We shall discuss some examples of such actions in more details further.

Since they are free, the quotients of ${\mathbb S}^3$ by these isometric actions constitute regular manifolds, locally isometric to ${\mathbb S}^3$. A convenient coordinatisation is thus provided by the Euler angles on ${\mathbb S}^3$, in terms of which
\ba
z_1 &=& \cos \frac{\theta}{2} e^{\frac{i}{2} \left (\phi + \psi \right ) } \, ,\\
z_2 &=& \sin \frac{\theta}{2} e^{\frac{i}{2} \left (\phi - \psi \right ) } \, ,
\ea
where $\theta \in [0, \pi)$, $\phi \in [0, 2\pi)$ and $\psi \in [0, 4\pi)$. The standard flat metric of ${\mathbb R}^4$ induces a maximally symmetric metric $h$ on the 3-sphere, which in the Euler angle coordinates reads
\ba
h &=& \frac{1}{4} \left [d\theta^2 + \sin^2 \theta d\phi^2 + \left ( d\psi + \cos \theta d \phi \right )^2\right ] \nonumber \\
&=& \frac{1}{4} \left [d\theta^2 + d\phi^2 + d\psi^2+ \cos \theta \left ( d\phi \otimes d\psi +d\psi \otimes d\phi \right ) \right ]
\ea
In these coordinates, the fibration of ${\mathbb S}^3$ as the fiber bundle ${\mathbb S}^1 \rightarrow {\mathbb S}^2$ is manifest.
According to the previous subsection, this metric on ${\mathbb S}^3$ induces a constant curvature metric on each of the quotients ${\mathbb S}^3/\Gamma$ mentioned here. In what follows, the riemannian 3-manifolds thus obtained will serve as the base spaces of conical 4-spaces. The well-known self-dual instanton metrics with asymptotic infinity ${\mathbb S}^3/\Gamma$ will provide regularising metrics for the associated cone metrics, allowing us to compute their distributional Euler form as the limit of a family of smooth forms. Alternatively, from the point of view of algebraic geometry, the conical 4-spaces obtained will just appear to be the well known quotient singularities ${\mathbb C}^2/\Gamma$ and we shall therefore use the standard results associated, as well as the distributional version of the Chern-Gauss-Bonnet theorem (\ref{DistCGB}), to compute the distributional Euler form of the most complicated examples. In the following subsections, we first concentrate on the single actions, providing the distributional interperetation of the Euler form in all cases. Then, we give a brief description of the calculations involved in the cases of double and linked actions.

\subsection{The case of ${\mathbb R}^4/{\mathbb Z}_2$}
In this subsection, we study the geometry of the quotient space ${\mathbb R}^4/{\mathbb Z}_2$, obtained by pairwise identification of the points $(x,y)$ of ${\mathbb R}^4$ such that $x=-y$. This equivalence is non-degenerate everywhere on ${\mathbb R}^4$ except at $0$, which is a fixed point of the action of ${\mathbb Z}_2$. Given a foliation of ${\mathbb R}^4$ by 3-spheres, we can deduce a foliation of the quotient space ${\mathbb R}^4/{\mathbb Z}_2$ by the real projective space ${\mathbb R}{\mathbb P}^3$, obtained from the 3-sphere by identification of its antipodal points, {\it i.e.} by single action of ${\mathbb Z}_2$. This quotient space is regularisable and a regularisation is provided by the well known Eguchi-Hanson metric \cite{EH}. This metric first appeared in the context of self-dual gravitational instantons \cite{EGH}, where it constitutes a one parameter family of regular finite energy solution of Einstein vacuum equations. As we shall see, for a certain value of its parameter, the Eguchi-Hanson metric degenerates to the cone metric on ${\mathbb R} \times {\mathbb R}{\mathbb P}^3$. We can therefore use this family of metrics to compute the Euler form and then prove that, in the limit, the latter tends to a distribution. For all $t \in {\mathbb R}$, the four dimensional Eguchi-Hanson metric is given by
\be
\label{EH4}
g_{EH} (t) = \frac{dr \otimes dr}{1-\frac{t^4}{r^4}} + r^2 \left (\sigma^x \otimes \sigma^x + \sigma^y \otimes \sigma^y \right ) + r^2 \left (1-\frac{t^4}{r^4} \right) \sigma^z \otimes \sigma^z \, ,
\ee
with the left invariant 1-forms of the group manifold $SU(2)$
\ba
\label{tx}
2\sigma^x &=& \sin (\psi )  d\theta - \sin(\theta) \cos(\psi) d\phi \\
\label{ty}
2\sigma^y &=& \cos (\psi) d\theta + \sin(\theta) \sin(\psi) d\phi \\
\label{tz}
2\sigma^z &=& d\psi + \cos(\theta) d\phi \, .
\ea
The Euler angles $(\theta,\phi,\psi)$ are usually taken within the ranges $\theta \in [0, \pi)$, $\phi \in [0, 2\pi )$ and $\psi \in [0, 4\pi )$ on ${\mathbb S}^3$ but here, in order to avoid a bolt singularity at $r=t$, we shall assume that $\psi \in [0, 2 \pi )$, in which case, we only have a coordinate singularity and ${\mathbb S}^3$ is reduced to ${\mathbb S}^3/{\mathbb Z}_2$. Notice that the metric (\ref{EH4}) is singular at $r=0$. Moreover for $r<t$, it is no longer riemannian since both $\partial_r$ and $e_z=\partial_\psi + \frac{1}{\cos(\theta)} \partial_\phi$ become timelike. However, we can get a regular riemannian metric provided that we restrict ourselves to the range $r \in [t, +\infty) $. For future convenience, we introduce the coordinate $\rho = r-t$. Notice that, for all $t \neq 0$, the topology of the resulting manifold is locally ${\mathbb R}^2 \times{\mathbb S}_2$ in the vicinity of $\rho =0$. On the other hand, for large values of $\rho$, we get asymptotically ${\mathbb R} \times {\mathbb S}_3/{\mathbb Z}_2 =  {\mathbb R} \times {\mathbb R} {\mathbb P}^3 $, i.e. the four dimensional euclidian space with antipodal points indentified. Indeed, the Eguchi-Hanson manifold $M_4$ can be seen as a fiber bundle ${\mathbb R}^2 \rightarrow {\mathbb S}^2$, where the fiber ${\mathbb R}^2$ is spanned by $(\rho, \psi)$ and the base manifold ${\mathbb S}^2$ is spanned by $(\theta, \phi)$. This bundle interpolates between ${\mathbb R} \times {\mathbb R} {\mathbb P}^3$ at infinity, where the $U(1)$ Hopf fiber generated by $\partial_\psi$ fibers ${\mathbb S}^2$ to yield the ${\mathbb R} {\mathbb P}^3$, and ${\mathbb R} \rightarrow {\mathbb S}^2$ at $\rho=0$, where the $U(1)$ fiber shrinks to zero and the size of the ${\mathbb S}^2$ is controlled by the value of $t$. In the limit, when $t$ tends to zero, the ${\mathbb S}^2$ at $\rho=0$ shrinks to a point and we recover the singular fixed point of ${\mathbb R}^4/{\mathbb Z}_2$. The Eguchi-Hanson manifold $M_4$ can thus be thought of as a resolution $\pi : M_4 \rightarrow {\mathbb R}^4/{\mathbb Z}_2$ of the ${\mathbb R}^4/{\mathbb Z}_2$ singularity, with exceptional divisor $\pi^{-1}(0_{{\mathbb R}^4/{\mathbb Z}_2}) = {\mathbb S}^2$, and the Eguchi-Hanson metrics (\ref{EH4}), for all non-vanishing values of $t$, constitute a familly of ${\mathcal C}^\infty(M_4)$ riemannian metrics over $M_4$, regularising the cone metric over ${\mathbb R}\times{\mathbb R} {\mathbb P}^3$ which is obtained for $t=0$. As in the previous subsection, we shall thus compute the Euler form of $g_{EH}(t)$ for all values of the real parameter $t$, and show explicitely that, in the limit, when $t$ tends to zero, the Euler form becomes a distribution.

For all $t$, an orthonormal basis for the Eguchi-Hanson metric (\ref{EH4}) is given by
\ba
\theta^r (t) &=& \frac{1}{\left (1-\frac{t^4}{r^4} \right )^{1/2}} \\
\theta^x &=& r \sigma^x \\
\theta^y &=& r \sigma^y \\
\theta^z (t) &=& r \left (1-\frac{t^4}{r^4} \right )^{1/2} \sigma^z \, .
\ea
Let us first introduce the connection 1-form $\omega^a{}_b(t)$ of $g_{EH}(t)$. Since the left invariant 1-forms of $SU(2)$ (\ref{tx}-\ref{tz}) satisfy the Cartan-Maurer structure equation
\be
d\sigma^i = \epsilon_{ijk} \sigma^j \wedge \sigma^k \, ,
\ee
we readily deduce the components of the metric and torsion free connection 1-form $\omega^a{}_b(t)$,
\ba
\label{omegaEH1}
\omega^x{}_r (t) =  \left (1-\frac{t^4}{r^4} \right )^{1/2} \sigma^x \qquad && \qquad \omega^x{}_y (t)=\left (1+\frac{t^4}{r^4} \right ) \sigma^z \\
\omega^y{}_r (t) =  \left (1-\frac{t^4}{r^4} \right )^{1/2} \sigma^y \qquad && \qquad \omega^z{}_x(t) = \left (1-\frac{t^4}{r^4} \right )^{1/2} \sigma^y \\
\label{omegaEH2}
\omega^z{}_r (t) = \left (1+\frac{t^4}{r^4} \right ) \sigma^z \qquad && \qquad \omega^y{}_z (t)= \left (1-\frac{t^4}{r^4} \right )^{1/2} \sigma^x \, .
\ea
For all $t$, the curvature 2-form of this connection, $\Omega^a{}_b(t) = d\omega^a{}_b(t) +\omega^a{}_c(t) \wedge \omega^c{}_b(t)$, is then given by 
\ba
\label{OmegaEH1}
\Omega^x{}_r(t) = - \frac{2 t^4}{r^6} \left (\theta^x \wedge \theta^r + \theta^y \wedge \theta^z \right )  \qquad && \qquad \Omega^x{}_y (t)= \frac{4 t^4}{r^6} \left (\theta^x \wedge \theta^y - \theta^r \wedge \theta^z \right ) \\
\Omega^y{}_r (t)= - \frac{2 t^4}{r^6} \left (\theta^y \wedge \theta^r + \theta^z \wedge \theta^x \right ) \qquad && \qquad \Omega^x{}_z (t)= - \frac{2 t^4}{r^6} \left (\theta^x \wedge \theta^z + \theta^r \wedge \theta^y \right ) \\
\label{OmegaEH2}
\Omega^z{}_r (t) = - \frac{4 t^4}{r^6} \left (\theta^z \wedge \theta^r + \theta^x \wedge \theta^y \right ) \qquad && \qquad \Omega^y{}_z(t) = - \frac{2 t^4}{r^6} \left (\theta^y \wedge \theta^z - \theta^r \wedge \theta^x \right ) \, .
\ea
Notice that this curvature 2-form is self-dual, which  further implies that the metric is a solution of Einstein's equations in the vacuum, {\it i.e.} for all $t$
\be
\Omega^{bc}(t) \wedge \theta^\star_{abc}(t) = 0 \, .
\ee

For all non-vanishing values of $t$, we get from (\ref{OmegaEH1}-\ref{OmegaEH2}), that
\ba
\label{EHEuler}
E_4(t) &=& \Omega^{ab}(t) \wedge \Omega^{cd}(t) \wedge \theta^\star_{abcd}(t) = 384 \frac{t^8}{(\rho+t)^{12}} \theta^\star(t) \nonumber \\
&=& 48 \frac{t^8}{(\rho+t)^{9}} \sin \theta \, d\rho \wedge d\theta \wedge d\phi \wedge d\psi\, .
\ea
We shall now prove that, in the limit when $t$ tends to zero, this 4-form admits a distributional limit. In order to do so, 
we now wish to consider the weighted integral of this quantity over an open subset $U \subset M_4$, corresponding to the points such that $0 \leq \rho \leq R$, when the parameter $t$ is small. Let therefore $f:U\rightarrow {\mathbb R}$ be a ${\mathcal C}^1(U)$ function with compact support in $U \subset M_4$.  For all $t \in (0,1]$, the weighted integral of $E_4(t)$ yields
\ba
\int_U f E_4(t) &=& 48 t^8 \int d\rho \, d\theta \, d\phi\, d\psi \, f(\rho,\theta,\phi,\psi) \frac{\sin(\theta)}{(\rho + t)^9} \nonumber \\
&=& 48 t^8 \int d\rho \, d\theta \, d\phi\, d\psi \, \left [\frac{f'}{8(\rho + t)^8} - \left ( \frac{f}{8(\rho +t)^8} \right )'  \right ] \sin(\theta) \nonumber \\
&=&48 t^8 \int d\rho \, d\theta \, d\phi\, d\psi \, \frac{f'}{8(\rho + t)^8} \sin(\theta) \nonumber \\
&&+ 48 \int d\theta \, d\phi\, d\psi \, \left (\frac{f(0,\theta,\phi,\psi)}{8} - t^8\frac{f(R,\theta,\phi,\psi) }{8(R+t)^8} \right ) \sin(\theta) \nonumber\\ 
&=& 48 \pi^2 f(0) + O(t) \, . 
\ea
In the limit, when $t$ tends to zero, the above equation admits a distributional interpretation. The latter consists in saying that there exists a distribution $\hat{E}_4$, such that for all ${\mathcal C}^1(U)$ function $f:U\rightarrow {\mathbb R}$ with compact support in $U \subset M_4$, 
\be
\left\langle \hat{E}_4 ; f \right\rangle_U = \lim_{t \to 0} \int_U f E_4(t) = \left \{ \begin{array}{cc}
48 \pi^2 f(0) & \mbox{if $0 \in U$;}\\
0 & \mbox{otherwise.}
\end{array} \right .
\ee
Up to a constant coefficient, this is precisely the definition of the $0$-centered Dirac distribution over $M_4$. We thus get 
\be
\label{distint}
\hat{E}_4 = 48 \pi^2 \delta_0^{(4)} \, ,
\ee
where, for simpicity, we have included the volume form $\theta^\star$ in the definition of the Dirac distribution $\delta_0^{(4)}$.

As a consequence of (\ref{distint}), the Chern-Gauss-Bonnet theorem for a compact subset $U \subset {\mathbb R}^4/{\mathbb Z}_2$, with boundary $\partial U$,
becomes
\be
\left\langle \hat{E}_4 ; 1 \right\rangle_U = 32 \pi^2 \chi(U) - \int_{\partial U} \Phi_4(0) \, ,
\ee
where
\ba
\Phi_4(0) &=& \lim_{t \to 0} \, \, \epsilon_{ijk} \left (\frac{8}{3} \omega^i{}_r (t) \wedge \omega^j{}_r(t) \wedge \omega^k{}_r(t)+ 4 \Omega^{ij}(t) \wedge \omega^k{}_r(t)  \right ) \nonumber \\
&=& \lim_{t\to 0} \, \, 2 \left (1+3\frac{t^8}{r^8}  \right )\sin \theta d\theta \wedge d\phi \wedge d\psi 
\ea
is the limit, when $t$ tends to zero, of the Chern-Simons 3-form associated to $E_4(t)$ and $\chi(U)=2$. Notice that the Chern-Gauss-Bonnet theorem also holds for the whole family of Eguchi-Hanson metrics, that is for all values of $t$. The Euler characteristic of the Eguchi-Hanson manifold could be computed otherwise, just using the knowledge we have of its topology. As already stated, it can be seen as a minimal resolution of the singularity ${\mathbb R}^4/{\mathbb Z}_2$ with exceptional divisor $\pi^{-1} (0_{{\mathbb R}^4/{\mathbb Z}_2}) = {\mathbb S}^2$. From the definition of a minimal resolution of singularity \cite{Slodowy}, we also know that ${\mathbb R}^4/{\mathbb Z}_2 \backslash \{0_{{\mathbb R}^4/{\mathbb Z}_2}\}$ is isomorphic to $\widehat{{\mathbb R}^4/{\mathbb Z}_2} \backslash \pi^{-1} (0_{{\mathbb R}^4/{\mathbb Z}_2})$ through $\pi$. The Euler characteristics of these regular manifolds must therefore be equal and we have
\be
\chi \left ( \widehat{{\mathbb R}^4/{\mathbb Z}_2} \right ) - \chi \left ({\mathbb S}^2 \right) = \chi \left ( {\mathbb R}^4/{\mathbb Z}_2 \backslash \{0_{{\mathbb R}^4/{\mathbb Z}_2}\} \right ) \, .
\ee
The r.h.s. obviously vanishes and, since $\chi \left ({\mathbb S}^2 \right)=2$, we obtain the desired result.
As a matter of fact, the value of the Euler characteristic of the Eguchi-Hanson manifold is just one plus the number of non-trivial irreducible representations, or equivalently of conjugacy classes, of ${\mathbb Z}_2$, {\it i.e.} $\chi(U) = 1+1$. As we shall see in the next subsection, this is just the simplest instance of the McKay correspondence for the Euler characteristic \cite{MK}. More generally, in algebraic geometry, the McKay correspondence relates the irreducible representations of a finite subgroup $\Gamma \subset SL(2, {\mathbb C})$ to the cohomology $H^*\left (\widehat{{\mathbb C}^2/\Gamma} \right )$ of the minimal resolution $\pi : \widehat{{\mathbb C}^2/\Gamma} \rightarrow {\mathbb C}^2/\Gamma $ of the singular quotient space ${\mathbb C}^2/\Gamma$. It thus provides a natural way to compute the Euler characteristic of the minimal resolutions of single action quotients. In the following examples, we shall use these results together with the distributional Chern-Gauss-Bonnet theorem (\ref{DistCGB}) as direct computations in the regularising metrics become more and more involved.

\subsection{The case of ${\mathbb R}^4/{\mathbb Z}_m$ and the $A_{m-1}$-series singularities}
The Eguchi-Hanson metric presented in the previous subsection, can be thought of as the regularisation of the singular quotient space ${\mathbb R}^4/{\mathbb Z}_2$. Actually, the latter is just the simplest case of an $A$-series singularity on ${\mathbb C}^2$. Higher order singularities in this series can be obtained by identifications under the general cyclic subgroups ${\mathbb Z}_m \subset SU(2)$. For all $m\geq 2$, the quotient ${\mathbb R}^4/{\mathbb Z}_m$ is singular at the fixed point $0 \in {\mathbb R}^4$ of the ${\mathbb Z}_m$-action, but it admits a minimal resolution of singularities. Its singular cone metric can also be regularised by a generalised version of the Eguchi-Hanson metric due to Gibbons and Hawking \cite{GH}, which reads
\ba
\label{GH4}
g_{GH}({\bf x}_i) &=& V^{-1} \left (dz - A \right )^2 + V d{\bf x} \cdot d {\bf x} \, , \\
V &=& \sum_{i=0}^{m-1} \frac{1}{| {\bf x} - {\bf x}_i |} \, ,
\ea
where $z \in [0, 4\pi)$, ${\bf x} \in {\mathbb R}^3$ and  the 1-form $A$ is given by $dV = \star_3 dA$. The Hodge dual is taken with respect to the euclidean 3-metric $d{\bf x} \cdot d{\bf x}$. As for the Eguchi-Hanson metric, the Gibbons-Hawking metric first appeared in the context of self-dual gravitational instantons as a regular finite energy multi-instanton solution of Einstein vacuum equations with ${\mathbb S}^3/{\mathbb Z}_m$ asymptotic infinity. It depends on a set of parameters encoded in the $m$ fixed vectors ${\bf x}_i$. It can be shown \cite{Prasad} that, if $m=2$, (\ref{GH4}) reduces to the Eguchi-Hanson metric (\ref{EH4}) in a different coordinate system. On the other hand, for all values of $m \geq 2$, we have a fiber bundle $U(1) \rightarrow {\mathbb R}^3$ whose fiber $U(1)$ shrinks to zero size at the points ${\bf x}_i \in {\mathbb R}^3$ of the base space. We thus have a series of $m-1$ topological 2-spheres with pointlike intersections at the ${\bf x}_i$. Their size is therefore controlled by the modules $|{\bf x}_i - {\bf x}_j|$. From the point of view of algebraic geometry, these correspond to the $(m-1)$ ${\mathbb{CP}}^1$ constituting the exceptional divisor associated to the minimal resolution of singularities of ${\mathbb C}^2/{\mathbb Z}_m$. For all $m \geq 2$, the configuration of these ${\mathbb{CP}}^1$ is given by the Dynkin diagram 
\be
\Delta (A_{m-1}) \quad  = \quad \circ - \circ - \cdots - \circ
\ee 
of the simply laced Lie algebra $A_{m-1}$, under the identification of each ${\mathbb{CP}}^1$ with a vertex of $\Delta (A_{m-1})$ and of each intersection point with the corresponding edge of $\Delta (A_{m-1})$ \cite{Slodowy}.

The calculations of the previous subsection can be repeated for the Gibbons-Hawking metrics (\ref{GH4}), to obtain
\be
\label{cyclicEuler}
\hat{E}_4 = 32 \pi^2 \left (m-\frac{1}{m} \right ) \delta_0^{(4)} \qquad \mbox{on ${\mathbb R}^4/{\mathbb Z}_m$.}
\ee
The Chern-Gauss-Bonnet theorem is also valid for all values of the ${\bf x}_i$ and in the limit, when they all collapse to ${\bf 0}$, yielding $\chi = m$. This result can also be obtained otherwise from standard results in algebraic geometry. Indeed, as already stated, ${\mathbb C}^2/{\mathbb Z}_m$ admits a minimal resolution of singularities by a regular manifold $\widehat{{\mathbb C}^2/{\mathbb Z}_m}$. The configuration of its exceptional divisor is given by $\Delta (A_{m-1})$. Along the lines of the previous subsection, the Euler characteristic of $\widehat{{\mathbb C}^2/{\mathbb Z}_m}$ is therefore obtained by adding a contribution of $2$ for each of the $m-1$ vertices of $\Delta (A_{m-1})$, corresponding to the $(m-1)$ ${\mathbb{CP}}^1$, and of $-1$ for each of its $m-2$ edges, corresponding to the $m-2$ intersection points of the $m-1$ ${\mathbb{CP}}^1$. As in the previous subsection, the
Euler characteristic of $\widehat{{\mathbb C}^2/{\mathbb Z}_m}$ is also related to the number of isomorphism classes of non-trivial irreducible representations of ${\mathbb Z}_m$ through the McKay correspondence \cite{MK}. Using either result, we thus have $\chi \left (\widehat{{\mathbb C}^2/{\mathbb Z}_m} \right ) =m$. On the other hand, the boundary integral is known, (\ref{boundary}), and we have
\be
\int_{{\mathbb S}^3/{\mathbb Z}_m} \Phi_4 = \frac{1}{|{\mathbb Z}_m|} \int_{{\mathbb S}^3} \Phi_4 = \frac{1}{m} \int_{{\mathbb S}^3} \Phi_4 = \frac{32 \pi^2}{m} \, .
\ee
The normalisation in (\ref{cyclicEuler}) thus follows.

\subsection{The $D_{m+2}$ and $E_{6,7,8}$-series singularities}
To complete the single action picture, we shall now discuss the $D$ and $E$-series singularities. These correspond to single action quotient spaces of the form ${\mathbb R}^4/\Gamma$, where $\Gamma$ is either $D^*_{m}$, $T^*$, $O^*$, or $I^*$ for the $D_{m+2}$, $E_6$, $E_7$ or $E_8$-series singularities respectively.
In these cases, although their existence was strongly motivated after the works of Calabi \cite{Calabi} and Hitchin \cite{Hitchin}, explicit knowledge of the corresponding self-dual instanton metrics was only achieved by Kronheimer in \cite{Kronheimer}, using hyper-Kh\"aler quotients techniques. These provide a set of regularising metrics for the cone metrics on the corresponding quotients. The calculations in the regularised metrics performed in the previous subsections could be extended to these regularised metrics as well. However, using the results of the previous subsections on the Euler characteristic of the resolution of singularities, we can argue that the Euler form of the quotient spaces ${\mathbb R}^4/\Gamma$, when $\Gamma$ is either $D^*_m$, $T^*$, $O^*$ or $I^*$, are also given by a Dirac distribution. All of these quotient spaces admit a minimal resolution of singularities whose exceptional divisor's configuration  is given by the Dynkin diagrams

 \ba
\begin{array}{rclcrcc}
 &\circ& & \qquad & & \circ & \nonumber \\
                           & | & & \qquad & & | & \nonumber \\
\Delta \left (D_{m+2} \right ) \quad = \quad \circ - \circ- \cdots - &\circ  & - \circ & \qquad \qquad &\Delta \left (E_{6} \right ) \quad = \quad \circ - \circ- &\circ  & - \circ - \circ \nonumber \\
&&&&&& \nonumber \\
&\circ& & \qquad & & \circ & \nonumber \\
                           & | & & \qquad & & | & \nonumber \\
\Delta \left (E_{7} \right ) \quad = \quad \circ - \circ - \circ- &\circ  & - \circ - \circ & \qquad & \Delta \left (E_{8} \right ) \quad = \quad \circ - \circ -\circ - \circ- &\circ  & - \circ - \circ
\end{array}
\ea
of the associated simply laced Lie algebras.
The normalisation of the distributional Euler form therefore follows from the distributional version of the Chern-Gauss-Bonnet theorem (\ref{DistCGB}), since the Euler characterstics of the regular spaces $\widehat{{\mathbb R}^4/\Gamma}$ can be obtained either by a direct computation on the Dynkin diagrams $\Delta \left ( \Gamma \right )$ or, through the McKay correspondence \cite{MK}, as the number of isomorphism classes of non-trivial irreducible representations of $\Gamma$. We thus have 
\be
\chi \left (\widehat{{\mathbb R}^4/D^*_{m}} \right ) = m+3 \, , \quad \;  \chi \left (\widehat{{\mathbb R}^4/T^*} \right )=7 \, ,
\quad \; \chi \left (\widehat{{\mathbb R}^4/O^*} \right )=8 \, ,  \quad \; \chi \left (\widehat{{\mathbb R}^4/I^*} \right )=9 .
\ee
On the other hand, the orders of the above finite subgroups are given by
\be |D^*_{m}| = 4m\, , \qquad \qquad |T^*| =24 \, , \qquad \qquad |O^*| =48 \, \qquad \mbox{and} \qquad |I^*| =120 \, , \ee
allowing us to evaluate the corresponding boundary integrals in (\ref{DistCGB}), by use of (\ref{boundary}). We therefore get
\be
\label{EulerDE}
\hat{E}_4 = \left \{ \begin{array}{cc}
                      32 \pi^2 \left (m+3 - \frac{1}{4m}\right )  \delta_0^{(4)}  & \qquad \mbox{on ${\mathbb R}^4/D^*_{m}$;} \\
                      \frac{668\pi^2}{3} \delta_0^{(4)} & \qquad \mbox{on ${\mathbb R}^4/T^*$;} \\
                      \frac{766\pi^2}{3} \delta_0^{(4)} & \qquad \mbox{on ${\mathbb R}^4/O^*$;} \\
                      \frac{4316 \pi^2}{15} \delta_0^{(4)} & \qquad \mbox{on ${\mathbb R}^4/I^*$.}
                     \end{array}
\right .
\ee
These relations, together with eq. (\ref{cyclicEuler}) constitute the distributional interpretation for the Euler forms of all the single action quotient spaces.

\subsection{Double and linked action singularities}
Before we proceed with the explicit construction of self-gravitating branes of codimension 4, let us now focus on the distributional interpretation of the double and linked action quotient spaces. The simplest example of such actions is obtained when both the right and left subgroups are cyclic, {\it i.e.} $\Gamma = {\mathbb Z}_m \times {\mathbb Z}_{n}$. The natural representation of ${\mathbb Z}_m$ on ${\mathbb C}^2$ is generated by
\be
\label{repZ2}
\zeta_m = \left (\begin{array}{cc}
       e^{i \frac{2 \pi}{m}} & 0 \\
       0    & e^{-i \frac{2\pi}{m}} \\
       \end{array}
 \right )  \in SU(2)
\ee
For all integer $n$, the linked action of ${\mathbb Z}_{2n} \times_{q} {\mathbb Z}_{2n}$ on $SU(2)$ generated by $(\zeta_{2n}^{1+n-q}, \zeta_{2n}^{1+n+q})$, where $0<q<n$ and gcd$(n,q)=1$, translates into an action on ${\mathbb C}^2$ of the cyclic subgroup of $U(2)$ generated by 
\be
\label{cnq}
C_{n,q} = \left (\begin{array}{cc}
       e^{i  \frac{2 \pi}{n} } & 0 \\
       0    & e^{i 2\pi \frac{q}{n}} \\
       \end{array}
 \right ) \in U(2)
\ee 
For all $n$, the quotient space ${\mathbb R}^4 /\left ({\mathbb Z}_{2n} \times_q {\mathbb Z}_{2n} \right )$ is therefore obtained as the quotient of ${\mathbb C}^2$ by the action of the cyclic group generated by $C_{n,q}$, which is free everywhere except at $0$. From the foliation of ${\mathbb R}^4$ by ${\mathbb S}^3$, we deduce that, for all $n$, ${\mathbb R}^4/\left ({\mathbb Z}_{2n} \times_q {\mathbb Z}_{2n} \right )$ admits a foliation by the lens space ${\mathbb L}(n,q) = {\mathbb S}^3/({\mathbb Z}_{2n} \times_q {\mathbb Z}_{2n})$. From the point of view of algebraic geometry, these quotient spaces are known to admit a minimal resolution whose dual graph is obtained through the Hirzebruch-J\"ung continued fraction, \cite{Brieskorn},
\be
\frac{n}{q} = b_1 - \frac{1}{b_2-\frac{1}{\begin{array}{ccc}
                                                  b_3 -  &  & \\
                                                         & \ddots & \\
                                                         &         & \frac{1}{b_{r-1}- \frac{1}{b_r}}
                                                 \end{array}
}} \equiv \left [b_1, b_2, \cdots, b_r \right ] \, ,
\ee
as 
\ba
\label{line}
\left \langle n,q \right \rangle \quad  &=& \quad \circ - \circ - \cdots - \circ \\
                                         && \;\;^{(b_1)} \;\; ^{(b_2)} \quad \quad \;\; ^{(b_r)} \nonumber
\ea
It has $r$ vertices corresponding to the $r$ ${\mathbb{CP}}^1$ in the decomposition of the exceptional divisor and $r-1$ edges corresponding to their intersection points. 
We thus get $\chi = 1+r$. Since the order of the linked action of ${\mathbb Z}_{2n} \times_q {\mathbb Z}_{2n}$ is $n$, the boudary term in (\ref{DistCGB}) is known from (\ref{boundary}). We thus get
\be
\label{cyclic2Euler}
\hat{E}_4 = 32 \pi^2 \left ( 1+r-\frac{1}{n} \right ) \delta^{(4)}_0
\ee
As an example, let us consider the case when $n=12$ and $q=5$. Then, we have
\be
\frac{12}{5} = 3-\frac{1}{2-\frac{1}{3}} = \left [3,2,3 \right ] \, .
\ee
The dual graph is therefore 
\ba
\left \langle 12,5 \right \rangle \quad  &=& \quad \circ - \circ - \circ \\
                                          && \;\;\, ^{(3)} \;\;\, ^{(2)} \;\;\, ^{(3)} \nonumber
\ea
with $r=3$ and we have $\chi = 4$. In this case, eq. (\ref{cyclic2Euler}) gives
\be
\label{l125}
\hat{E}_4 = \frac{376}{3} \pi^2 \delta_0^{(4)} \, , \qquad \mbox{on ${\mathbb R} \times {\mathbb L}(12,5)$.}
\ee
Notice also that for $q=n-1$ in (\ref{cnq}), we recover the single left action of ${\mathbb Z}_n$ and the dual graphs $\left \langle n,n-1 \right \rangle$ are precisely the Dynkin diagrams of the $A_{n-1}$-series. The corresponding $A$-series quotient spaces therefore admit a foliation by the lens spaces ${\mathbb L}(n,n-1)$.

As happens for the linked action of ${\mathbb Z}_{2n} \times_q {\mathbb Z}_{2n}$, all the double and linked actions mentioned in the subsection on spherical 3-manifolds, induce the action of a finite subgroup $G \subset GL(2, {\mathbb C})$ on ${\mathbb C}^2$ and the quotient space ${\mathbb C}^2/G$ admits a foliation by ${\mathbb S}^3/G$. Indeed, as stated in \cite{Seifert2}, the set of three dimensional manifolds foliating the quotient spaces ${\mathbb C}^2/G$, where $G$ is some finite subgroup of $GL(2, {\mathbb C})$, is in one-to-one correspondence with the spherical geometries ${\mathbb S}^3/\Gamma$, where $\Gamma$ is some finite subgroup of the isometry group $SO(4)$. The two pictures are therefore equivalent, but the ${\mathbb C}^2$ point of view is more suited for the study of minimal resolutions of singularities. The finite subgroups of $GL(2,{\mathbb C})$ were classified in \cite{Brieskorn}. They are all of the form
\ba
\label{H1H2}
(H_1, N_1 ; H_2, N_2) &\equiv & \left \{h \in GL(2,{\mathbb C}); \; h = h_1 \, h_2 \; \mbox{for some} \; h_1 \in H_1, h_2 \in H_2, \right . \nonumber \\
& &  \left . \qquad \qquad \qquad \quad  \mbox{such that} \; h_1 \; \mbox{mod} \; N_1 \; = \; h_2 \; \mbox{mod} \; N_2 \right \} \, ,
\ea
where $H_1$ stands for some finite subgroup of $ZL(2, {\mathbb C})$, the centre of $GL(2, {\mathbb C})$ and $H_2$ for some finite subgroup of $SL(2,{\mathbb C})$. $N_1$ and $N_2$ denote normal subgroups of $H_1$ and $H_2$ respectively. Obviously, $H_1$ can only be one of the cyclic subgroups $Z_k$ of $ZL(2, {\mathbb C})$ whose  natural representation on ${\mathbb C}^2$ is generated by
\be
e^{i \frac{2 \pi}{k} } \, \left ( \begin{array}{cc}
                                   1 & 0 \\
                                   0 & 1
                                  \end{array}
\right ) \, .
\ee
Notice that $Z_k$ should not be mistaken with ${\mathbb Z}_k \subset SU(2)$ whose natural representations on ${\mathbb C}^2$ is given by (\ref{repZ2}). From this point of view, the cyclic subgroups generated by the $C_{n,q}$ of eq. (\ref{cnq}) are obtained when both $H_1$ and $H_2$ are cyclic. The only other possibility will be to match the cyclic $H_1$ with one of the binary polyhedral groups $T^*$, $O^*$ or $I^*$. When $H_2$ is either $D_{n}^*$ or $T^*$, we shall distinguish the non-trivial normal subgroups ${\mathbb Z}_{2n} \lhd D_{n}^*$ and $D_{2}^* \lhd T^*$. All singular spaces obtained by taking the quotient of ${\mathbb C}^2$ under the action of a finite subgroup of the form of (\ref{H1H2}) admit a minimal resolution of singularities whose dual graph is of the form 
\ba
&\left \langle n_2, q_2 \right \rangle & \nonumber \\
&|& \nonumber \\
\left \langle b ; n_1, q_1; n_2, q_2; n_3, q_3 \right \rangle = &\left \langle n_1, q_1 \right \rangle - \circ - \left \langle n_3, q_3 \right \rangle& \nonumber \\
&(b)& \nonumber
\ea
where each $\left \langle n_i, q_i \right \rangle$ is a linear diagram like in (\ref{line}), with $0<q_i<n_i$ and $b\geq2$. We give, in the following table, the dual graphs $\Delta \left (G \right )$ corresponding to the different subgroups $G$ of $GL(2,{\mathbb C})$.

\vskip .5cm
\begin{center}
\begin{tabular}{|c|c|c|}
\hline
$G$ & Constraints & $\Delta \left (G \right )$ \\
\hline
$C_{n,q}$ & $0<q<n$ & $\left \langle n,q \right \rangle$ \\
           & gcd$(n,q)=1$ &  \\
\hline
$(Z_{2m}, Z_{2m}; D^*_{n}, D^*_{n})$ & $m=(b-1)n-q $ & $\left \langle b; 2,1; 2,1; n,q \right \rangle$ \\
                                     & gcd$(m,2)=1$ & \\
                                     & gcd$(m,n)=1$ & \\
&&\\
$(Z_{4m}, Z_{2m}; D^*_{n}, {\mathbb Z}_{2n})$ & $m =(b-1)n-q$  & $\left \langle b; 2,1; 2,1; n,q \right \rangle$ \\
                                              & gcd$(m,2)=2$  & \\
                                              & gcd$(m,n)=1$ & \\
\hline
$(Z_{2m}, Z_{2m}; T^*, T^*)$ & $m =6(b-2)+1$  & $\left \langle b; 2,1; 3,2; 3,2 \right \rangle$ \\
                             & $m=6(b-2)+5$   & $\left \langle b; 2,1; 3,1; 3,1 \right \rangle$ \\
&&\\
$(Z_{6m}, Z_{2m}; T^*, D^*_2)$ & $m =6(b-2)+3$  & $\left \langle b; 2,1; 3,1; 3,2 \right \rangle$ \\
\hline
$(Z_{2m}, Z_{2m}; O^*, O^*)$ & $m =12(b-2)+1$  & $\left \langle b; 2,1; 3,2; 4,3 \right \rangle$ \\
& $m =12(b-2)+5$  & $\left \langle b; 2,1; 3,1; 4,3 \right \rangle$ \\
& $m =12(b-2)+7$  & $\left \langle b; 2,1; 3,2; 4,1 \right \rangle$ \\
& $m =12(b-2)+11$  & $\left \langle b; 2,1; 3,1; 4,1 \right \rangle$ \\
\hline
$(Z_{2m}, Z_{2m}; I^*, I^*)$ & $m =30(b-2)+1$  & $\left \langle b; 2,1; 3,2; 5,4 \right \rangle$ \\
& $m =30(b-2)+7$  & $\left \langle b; 2,1; 3,2; 5,3 \right \rangle$ \\
& $m =30(b-2)+11$  & $\left \langle b; 2,1; 3,1; 5,4 \right \rangle$ \\
& $m =30(b-2)+13$  & $\left \langle b; 2,1; 3,2; 5,2 \right \rangle$ \\
& $m =30(b-2)+17$  & $\left \langle b; 2,1; 3,1; 5,3 \right \rangle$ \\
& $m =30(b-2)+19$  & $\left \langle b; 2,1; 3,2; 5,1 \right \rangle$ \\
& $m =30(b-2)+23$  & $\left \langle b; 2,1; 3,1; 5,2 \right \rangle$ \\
& $m =30(b-2)+29$  & $\left \langle b; 2,1; 3,1; 5,1 \right \rangle$ \\
\hline
\end{tabular}
\end{center}
\vskip .5cm

\noindent As we have seen in the previous subsections, the dual graphs provide an efficient way to determine the Euler characteristic of $\widehat{{\mathbb C}^2/G}$, for all $G$. Let us consider, for example, the quotient ${\mathbb C}^2/G$, where $G=(Z_{2}, Z_{2}; D^*_{4}, D^*_{4})$. From the above table, its dual diagram is simply
\ba
&\circ & (2) \nonumber \\
&|& \nonumber \\
\left \langle 2 ; 2, 1; 2, 1; 4, 3 \right \rangle = (2) \circ - &\circ & - \circ - \circ - \circ  \nonumber \\
&(2)& \;\;\,(2) \;\; (2) \;\, (2) \nonumber
\ea
The Euler characterisitic of its minimal resolution of singularities is therefore $\chi = 7$. On the other hand, the order of $(Z_{2}, Z_{2}; D^*_{4}, D^*_{4})$ is equal to $32$. According to (\ref{DistCGB}), we thus get
\be
\label{z2d4}
\hat{E}_4 = 223 \pi^2 \delta^{(4)}_{0}
\ee
Analogously, we can obtain the distributional Euler forms of all the double and linked action quotient spaces.

We shall now use the distributional interpretation of the Euler forms of these quotient spaces to construct self-gravitating branes solutions with energy-momentum tensors of the form of (\ref{set}). Notice, to conclude this section, that the distributional interpretation of the Euler forms of quotient spaces admitting a minimal resolution may be extended to other forms in $\Omega^*(M/\Gamma)$ and that its role in the context of the cohomology theory of these spaces probably deserves further attention.

\section{Codimension four (a)dS 3-branes in adS$_8$}
Although the construction of flat 3-branes in a flat eight-dimensional space-time would be straightforward, we directly construct an anti-de Sitter version of this problem since curved backgrounds are somehow more natural in the context of Lovelock theory. We should however bear in mind that the following results can be slightly modified to describe the flat case.

\subsection{The ${\mathbb Z}_2$-symmetric case}
We now assume that space-time is eight-dimensional and has the topology $M_8 = \Sigma_4 \times  {\mathbb R} \times {\mathbb R}{\mathbb P}^3$, where $\Sigma_4 
\hookrightarrow M_8$ is an embedded submanifold which we shall identify with the four dimensional standard universe. We endow this space-time with the warped metric $g_8(0)$ obtained as the limit when $t$ tends to zero of
\be
\label{3brane}
g_8 (t) = -k^2 r^2 d\tau^2+ \frac{dr^2}{k^2r^2} +r^2 \left [du^2+dv^2+ g_{EH}(t) \right ] \, ,
\ee
where, for all values of the real parameter $t$, $g_{EH}(t)$ denotes the four-dimensional Eguchi-Hanson metric. The connection 1-form associated to this metric is given by
\ba
\omega^\mu{}_\nu &=& \bar{\omega}^\mu{}_\nu \\
\omega^i{}_j(t) &=& \bar{\omega}^i{}_j(t) \\
\omega^i{}_r(t) &=& kr \bar{\theta}^i(t) \, ,
\ea
where we denote by greek or latin indices quantities refering to the coordinates $(\tau, r, u, v)$ or $(\rho, x, y, z)$ respectively and $\bar{\omega}^\mu{}_\nu$ and $\bar{\omega}^i{}_j$ respectively denote the components of the torsionfree metric connection 1-forms of the first fundamental forms of $\Sigma_4$ and $\Sigma_4^\perp$.

In turn, the curvature 2-form is given by
\ba
\Omega^\mu{}_\nu &=& -k^2 \theta^\mu \wedge \theta^\nu \\
\Omega^i{}_j(t) &=& \bar{\Omega}^i{}_j(t) - k^2 \theta^i(t) \wedge \theta^j(t) \\
\Omega^i{}_r(t) &=& -k^2 \theta^i(t) \wedge \theta^r(t) \, ,
\ea
where $\bar{\Omega}^i{}_j(t)$ is the induced curvature 2-form of $\Sigma_4^\perp$, {\it i.e.} the one associated to the Eguchi-Hanson metric (\ref{OmegaEH1}-\ref{OmegaEH2}). The latter being Ricci flat, {\it i.e.}
\be
\bar{\E}_{(1)i}(t) = \bar{\Omega}^{jk}(t) \wedge \theta^{\star}_{ijk}(t) =0 \, ,
\ee
we easily deduce that the l.h.s. of the field equations (\ref{love}) reduce to
\be
\left ( \alpha_0 - 42 \alpha_1 k^2 + 840 \alpha_2 k^4 -5040 \alpha_3 k^6 \right ) \theta_a^\star(t)  + \left (\alpha_2  - 18 \alpha_3 k^2 \right ) \bar{E}_{4}(t) \wedge \delta_a{}^\nu \theta^\star_{\Sigma \, \nu} \, ,
\ee
where $\bar{E}_4(t)$ is the Euler form of the Eguch-Hanson metric (\ref{EHEuler}) and $\theta^\star_{\Sigma \, \nu}$ is the Hodge dual of the standard basis of the $\theta^\mu$ on $\Omega^{(1)}(T\Sigma)$. In the limit when $t$ goes to zero, the distributional interpretation of the Euler class of the normal bundle of $\Sigma_4$, (\ref{distint}), therefore leads to the following regular field equations
\be
\label{FEq}
\Lambda + 21 k^2 - 420 \alpha_2 k^4 + 2520 \alpha_3 k^6 =0 \, ,
\ee
supplemented by a distributional part corresponding to the matching conditions
\be
\label{MC}
S^\mu{}_\nu = \left (-24 \pi ^2 \alpha_2 +  432 \pi^2 \alpha_3 k^2 \right ) \delta^\mu{}_\nu = \sigma \delta^\mu{}_\nu\, ,
\ee
where $\sigma$ is the tension of the 3-brane located at $\rho=0$, whose energy-momentum tensor is of the form of (\ref{set}). From the field equations (\ref{FEq}), we see that, in general, the constant curvature $k$ of $M_8$ is not uniquely determined, but rather belongs to the set of roots of a third order polynomial. On the other hand, in the matching conditions (\ref{MC}), notice the contribution of the induced curvature of $\Sigma_4$ through its Einstein tensor ${\mathcal E}^\Sigma_{(1) \mu} = -6k^2 \theta^\star_{\Sigma \mu}$, coming from the reduction of the Lovelock term of order three. Actually, the matching conditions (\ref{MC}) are just equivalent to the induced Einstein equations on the brane, with a cosmological constant $\Lambda_4 = -24 \pi ^2 \alpha_2$. This is a typical result for even codimension branes, as already mentioned in \cite{CZ1,CZ2}. We can furhter check that the above result (\ref{MC}) is consistent with the modified Lovelock couplings (\ref{alphatilde}) obtained there. Indeed, in this case, we have $\chi =2$ and the solid angle deficit in the transverse space is measured by
\be
\beta = \frac{1}{32 \pi^2} \int_{{\mathbb S}^3} \Phi_4 = \frac{1}{2} \, .
\ee
It is also worth emphasizing that here, unlike what happens in the case of the cosmic string, the value of $\beta$ seems rigidly fixed by the choice of the topology of the transverse space. This is actually a general result in the context of codimension four branes.

\subsection{A discrete spectrum for the tension}
For more general base spaces, the Eguchi-Hanson metric should be replaced in (\ref{3brane}) by the corresponding self-dual instanton metrics and, in the limit, we get a discrete spectrum for the tension of the brane, the tension increasing as the volume of the transverse space decreases. According to the results of the previous section for single action transverse spaces, namely (\ref{cyclicEuler}) and (\ref{EulerDE}), we have
\be
\sigma = \left \{ \begin{array}{cc} 
16\pi^2 \left ( m- \frac{1}{m} \right ) \left ( -\alpha_2  + 18 \alpha_3 k^2\right ) \, , \qquad & \qquad \mbox{for $\Sigma_4 \times {\mathbb R}^4/{\mathbb Z}_m$;} \\
16\pi^2 \left (m+3 - \frac{1}{4m}\right ) \left ( -\alpha_2  + 18 \alpha_3 k^2\right ) \, , \qquad & \qquad \mbox{for $\Sigma_4 \times {\mathbb R}^4/D^*_{m}$;} \\
\frac{334\pi^2}{3} \left ( -\alpha_2  + 18 \alpha_3 k^2\right ) \, , \qquad & \qquad \mbox{for $\Sigma_4 \times {\mathbb R}^4/T^*$;} \\
\frac{383 \pi^2}{3} \left ( -\alpha_2  + 18 \alpha_3 k^2\right ) \, , \qquad & \qquad \mbox{for $\Sigma_4 \times {\mathbb R}^4/O^*$;} \\
\frac{2158 \pi^2}{15} \left ( -\alpha_2  + 18 \alpha_3 k^2\right ) \, , \qquad & \qquad \mbox{for $\Sigma_4 \times {\mathbb R}^4/I^*$.} \\
                  \end{array}
\right .
\ee
We can also include the results from eq. (\ref{l125}) and (\ref{z2d4}) obtained as examples in the subsection on double and linked actions,
\be
\sigma = \left \{ \begin{array}{cc} 
\frac{188\pi^2}{3} \left ( -\alpha_2  + 18 \alpha_3 k^2\right ) \, , \qquad & \qquad \mbox{for $\Sigma_4 \times {\mathbb R} \times {\mathbb L}(12,5)$;} \\
\frac{223 \pi^2}{2} \left ( -\alpha_2  + 18 \alpha_3 k^2\right ) \, , \qquad & \qquad \mbox{for $\Sigma_4 \times {\mathbb R}^4/(Z_{2}, Z_{2}; D^*_{4}, D^*_{4})$.} \\
                  \end{array}
\right .
\ee
Notice that these energy levels are non-degenerate and that each value of the tension therefore corresponds to a unique topology. We could analogously derive the tensions corresponding to all possible topologies of the transverse space. This yields a complete spectrum of tensions corresponding to non-singular adS 3-branes in adS$_8$.
\\
\\

We have thus provided a distributional interpretation of the Euler forms of four dimensional quotient spaces admitting a minimal resolution of singularities. We derived it, either by regularising the corresponding singular conical metrics by self-dual instanton metrics with the chosen asymptotics, or by standard techniques from algebraic geometry. This has allowed us to construct a family of solutions of Lovelock equations describing localised adS 3-branes in adS$_8$. Like what happens in the case of a cosmic string, the matching conditions enforce the energy-momentum tensor of the brane to be tension-like and the geometry of the transverse space is obtained as the quotient of the flat space ${\mathbb R}^4$ by some discrete isometry group of $SO(4)$. From a physical viewpoint, the fact that the local geometry remains the same in the presence of the brane means that they have no gravitational interaction with other gravitating objects. However, unlike the cosmic string, the tension of these branes is not only related to the deficit angle of the conical transverse space, but also to their induced curvature and, more particularly, to their induced Einstein tensor. Moreover, the values of the tension belong to a discrete spectrum, parametrised by rational numbers in one-to-one correspondence with the finite subgroups of $SO(4)$, rather than to a continuous spectrum, parametrised by real numbers, as would be the case in codimension 2. The gap between two levels therefore measures the energy cost of a transition from one topology to another. The existence of such gaps and the absence so far of non-singular solutions with these symmetries may constitute indications that these solutions are stable and somehow unique in this class of symmetries. However, this remains to be proven. For that purpose, as well as to determine whether four dimensional gravity is localised on such 3-branes, the analysis of the linear stability of the solutions proposed here should provide valuable informations. In particular, although the moduli of the minimal resolution spaces usually appear as zero modes in the linear perturbations of the associated self-dual instantons, it is not clear what role they will play in the distributional limit of metrics of the form of (\ref{3brane}).

Notice, to conclude, that it should be possible to extend the construction presented in this paper to higher even codimensions, at least in the cases where the McKay correspondence has been established \cite{MK}. For example, the quotient space ${\mathbb C}^n/\Gamma$, where $\Gamma$ is the finite subgroup of $SU(n)$ generetad by the matrix diag$(z, \dots, z)$, with $z$ a primitive $n$-th root of unity, is known to admit a resolution of singularities with exceptional divisor reduced to a single ${\mathbb{CP}}^{n-1}$ and Euler characteristic $\chi = n$, \cite{MK}. This example of codimension $2n$ corresponds, in the codimension four context, to the simple ${\mathbb Z}_2$-symmetric case presented above. However it is very likely that more sophisticated examples can be used to construct solutions of higher codimension, with the same features as the ones presented here, {\it i.e.} contributions to the matching conditions of the induced curvature through reduced Lovelock terms of higher order and a discrete spectrum for the tension.
\vskip .85cm 
\noindent \emph{Acknowledgements} : It is a pleasure to thank Christos Charmousis for his useful comments on this manuscript. I
am also grateful to the EPSRC for financial support.

\end{document}